\documentclass[reprint, aip]{revtex4-1}

\usepackage{graphicx}
\usepackage{color}
\usepackage{booktabs}
\usepackage{amsmath}
\usepackage{multirow}
\usepackage{array}

\newcommand{\sci}[2]{\ensuremath{#1\times 10^{#2}}}
\newcommand{\sub}[1]{\ensuremath{_{\rm #1}}} 
\newcommand{\super}[1]{\ensuremath{^{\rm #1}}} 

\begin{document}

\title{Universal iso-density polarizable continuum model for molecular solvents}

\author{Deniz Gunceler}
\email[]{dg544@cornell.edu}
\author{T. A. Arias}
\affiliation{Department of Physics, Cornell University, Ithaca, New York, 14853, USA}

\keywords{Density Functional Theory, Quantum Chemistry, Polarizable Continuum Model}

\begin{abstract}
Implicit electron-density solvation models based on joint
density-functional theory offer a computationally efficient solution to the
problem of calculating thermodynamic quantities of solvated systems 
from first-principles quantum mechanics.
However, despite much recent interest in such models, to date the applicability of such models in the plane-wave context
to non-aqueous solvents has been limited
because the determination of the model parameters requires fitting to a large database of experimental solvation energies for
each new solvent considered.  This work presents an alternate
approach which allows development of new iso-density models for a large
class of protic and aprotic solvents from only simple, single-molecule {\em ab initio} calculations and
readily available bulk thermodynamic data.
\end{abstract}

\maketitle

\newcommand{\nr}[0]{\ensuremath{n(\vec{r})}}
\newcommand{\shape}[0]{\ensuremath{s(\nr)}}

{\em Ab initio} methods, in particular density-functional theory,
have a well established record of significant impact in chemistry, physics and materials science \cite{burke-perspective}.
However, despite the fact that many chemical reactions occur in a liquid environment
where the solvent (or electrolyte) plays an important role in the chemistry,
applications of density-functional methods to liquid and solvation chemistry
has lagged behind applications to solids, gas-phase molecules, or surfaces in vacuum.
This is because a single configuration of liquid molecules is often not representative of the thermal average, and
thermodynamic sampling, done for instance with {\em ab initio} molecular-dynamics\cite{CPMD} or QM/MM\cite{QMMM}, 
is needed to carry out realistic calculations.

Polarizable continuum models (PCMs) 
\cite{Tomasi94, Tomasi-Review, SMx96, SMxReview, SMD, PCM-Gygi, PCM-Marzari, PCM-Kendra, NonlinearPCM, PCM-Marzari-charged, SGA13} 
are one class of approximations where the solvent effect is reproduced
with an effective dielectric description of the liquid environment,
thus removing the need for sampling and thermodynamic integration and providing an economic alternative to molecular dynamics.
Continuum solvation models that use the union of spheres approach to construct the solute cavity,
in particular the universal "SMx" series\cite{SMx96, SMxReview, SMD} and those developed by Tomasi and coworkers,\cite{Tomasi94, Tomasi-Review}
have been very successful for a wide variety of solvents and widely implemented in quantum chemistry software.
An alternative approach is the iso-density PCM, \cite{PCM-Gygi, PCM-Marzari, PCM-Kendra, NonlinearPCM, PCM-Marzari-charged, SGA13} 
which is usually preferred in the plane-wave community and
uses the solute electron density to construct the solute cavity.
They also have achieved significant success describing processes in aqueous environments, including predicting, among other things, 
solvation energies for molecules \cite{NonlinearPCM, PCM-Marzari} and ions \cite{PCM-Marzari-charged, SGA13},
optical spectra for solvated molecules \cite{Marzari-excited}
as well as interfacial capacitances and potentials of zero charge for crystalline metals.\cite{PCM-Kendra}
Successful application of iso-density methods have great potential for new discoveries in many areas of research,
especially in energy-material related technologies\cite{Ceder-Li} where the processes at solid-liquid interfaces
are of prime importance and can most easily be studied in a plane-wave context.

Despite much recent interest, \cite{PCM-Marzari, PCM-Marzari-charged, chargeball-anatase}
one important obstacle is that most iso-density PCMs 
\cite{PCM-Sahak, PCM-Marzari, PCM-Gygi, chargeball-anatase, PCM-Marzari-charged}
are parametrized for only a handful of solvents (primarily water), 
and parameters don't exist for many solvents commonly encountered in organic chemistry and electrochemistry.
There has been much progress in reducing empiricism and increasing the
generality of such models\cite{PCM-Marzari, NonlinearPCM}, but such approaches continue to require
that multiple solvent-dependent parameters be fit to experimental data sets, 
typically to solvation free-energies of molecules.

While a great deal of solvation data is available 
\cite{truhlar, water-solvation-energies, water-solvation-energies-2, solvation-energies-3} 
for common solvents (such as water, chloroform and carbon tetrachloride),
many solvents of technological relevance do not have sufficient published data from which to construct iso-density continuum models.  
To aid exploration of microscopic physical processes in general solvent environments 
{\em and particularly studies to identify to best solvent for a given application}, 
this work provides a general framework for constructing accurate iso-density models for a large class of solvents,
similar in spirit to the way SMx models\cite{SMx96, SMxReview, SMD} have been made universal.
We use a limited number of coefficients, all of which can be obtained
directly \emph{either} from bulk thermodynamic data that is generally easy to
obtain \emph{or} from relatively simple single-molecule {\em ab initio} calculations.

{\em Universal solvation model ---} 
Polarizable continuum models represent a class of approximate theories which treat the interaction with the fluid environment as the dielectric response of a 
continuum medium filling the space not occupied by the molecule or surface of interest, to which we refer hereafter generically as the ``solute''.  
For most solvents, the dielectric response of the fluid is the largest, but not necessarily the entire, contribution to the free energy of solvation. 
Many models \cite{PCM-Gygi,PCM-Marzari,PCM-Kendra} assume that the dielectric response of the fluid is linear; but this need not be so.
Indeed, in this work, we use the nonlinear dielectric response function of Gunceler et. al. \cite{NonlinearPCM}, which also includes the rotational dielectric saturation of polar solvents.
Continuum solvation models generally model the remaining, non-electrostatic contributions to the free energy 
as related to the total area of the solute-solvent interface, treated independently of the underlying dielectric response model.
The central ideas in this paper thus apply equally well to linear response models, should one desire to work with those instead.

The first key issue in development of a PCM is determination of the dielectric region.
Traditionally, continuum models filled space with the dielectric medium except for spherical cavities centered on each atom of the solute,
each with a species-dependent atomic radius ultimately fit to a database of solvation energies\cite{Tomasi-Review}.
Several groups, instead, independently developed the isodensity approach for cavity determination \cite{PCM-Gygi,PCM-Sahak,chem-isodensity},
where the dielectric function changes from 1 in the interior of the solute region to the bulk value ($\varepsilon\sub{bulk}$) of the solvent dielectric constant, with the trasition occurring 
on the surface of a critical cutoff $n_c$ of the solute electron density.  As noted by Petrosyan {\em et al.}\cite{PCM-Sahak}, this has the advantage of placing such models in the class of approximate joint density-functionals.

For the functional form of the above transition in the the dielectric response, we here use the functional form of Petrosyan {\em et al.} \cite{PCM-Sahak},

\begin{equation}
    \varepsilon(\vec{r}) = 1 + (\varepsilon\sub{b} - 1) \hskip3pt \shape ,
\end{equation}
where \shape is the cavity-shape function, 
\begin{equation}
    \shape = \frac{1}{2}\textrm{erfc} \frac{\log(\nr/n_c)}{\sigma\sqrt{2}}.
\end{equation}
The parameter $\sigma$, controlling the width of the transition, is chosen to be large enough to resolve the transition on typical real-space grids. (Here, we employ $\sigma=0.6$ as chosen by Petrosyan and cowokers\cite{PCM-Sahak}.)
By replacing atom-dependent fit parameters with a single critical electron density $n_c$, which can then be fit to a database of solvation energies for each solvent considered, such iso-density approaches\cite{PCM-Gygi,PCM-Sahak,chem-isodensity} thereby eliminate many fit parameters in favor of a single parameter.  
However, the remaining cutoff parameter $n_c$ is highly solvent dependent, varying over several orders of magnitude, and it's determination still requires access to a database of solvation energies for each new solvent considered.

Even beyond the fitting needed to determine $n_c$, additional key parameters must be determined to yield accurate solvation energies.  This is because significant non-electrostatic processes contribute to solvation, such as the dispersion interaction between the solute and solvent, as well as the free-energy associated with forming the cavity in the solvent.  This is particularly true for non-aqueous solvents like chlorform and carbon tetrachloride, where solvation energies are not dominated by electrostatic interactions\cite{PCM-Tomasi-ChiralMolecules}.
To capture these effects, the effective surface-tension approximation is
commonly used in the iso-density PCM context. \cite{PCM-Marzari, NonlinearPCM} 
It approximates the \emph{non-electrostatic} contributions to the solvation energy ($E_{ne}$) as
\begin{equation} E_{ne} = \tau\sub{eff} \int d^3r \hskip3pt | \vec{\nabla}s | , \end{equation}
where $\tau\sub{eff}$ is an effective surface tension and the integral represents the surface area of the solute.
This particular use of the shape function to calculate the surface area
is a special case of the co-area formula in geometric measure theory. \cite{coarea}
For solvents with high bulk surface tension (such as water), the \emph{effective} surface tension generally is positive, whereas, for a large number of non-polar and weakly polar solvents with weak bulk surface tensions and strong (attractive) dispersion interactions, these {\em effective} surface tensions can become negative.

Motivated by the observation of a consistent trend of the effective surface tension with the strength of dispersion interactions, we now consider whether there exists a simple, approximate universal correlation between these two quantities.
To begin, we separate out the bulk surface tension, as suggested independently by Dupont and coworkers \cite{PCM-Marzari-charged},
and write $\tau\sub{eff} = \tau\sub{bulk} + \tau'$,
where $\tau\sub{bulk}$ is the (generally available) bulk surface tension of the solvent
and $\tau'$ is a correction term, which we will now attempt to correlate with dispersion interactions.  
Next, to estimate the strength of the dispersion interactions,
we make use of a very simple model and consider a self-solvation scenario where the van der Walls $r^{-6}$ potential
has been integrated in a region outside twice the van der Waals radius of the solvent molecule, resulting in
a dispersion energy per unit area $E\sub{vdw}/A \equiv \tau\sub{vdw}$ of
\begin{eqnarray}
    \tau\sub{vdw} & = & \frac{s_6}{A} \int_{2R_{vdw}}^{\infty} \hskip-10pt 4 \pi r^2 dr N\sub{b} \frac{ \sum_j C_6^{(j)} }{r^6} = \gamma\sub{1}  \frac{N\sub{b} C_{solv} }{R_{vdw}^5} \label{vdw}
\end{eqnarray}
where we employ the pair-potential model of dispersion corrections introduced by Grimme\cite{Grimme}.  
Here, $C_{solv} = \sum_j C_6^{(j)}$ is the effective dispersion coefficient, 
and is computed by summing over the Grimme $C_6$ coefficients of all atoms in the solvent molecule.
$N\sub{b}$ is the bulk number density of the solvent,
$R\sub{vdw}$ is a measure for the size of the solvent molecule (explained more detailed in the next paragraph) and
$s_6$ is a dimensionless scale factor accounting for renormalization of the fluctuating dipole interaction by multiple-atom interactions.  
In the second line of equation \ref{vdw}, we absorb
$s_6$ and all other dimensionless constants into $\gamma\sub{1}$.  (See below for a more detailed exploration of $s_6$.)
Finally, if desired, one can view the final expression as a simple dimensional analysis 
requiring some characteristic size of the solvent molecule, which we take to be the van der Waals radius.

\begin{center}
\begin{figure}
\center{\includegraphics[width=1.\columnwidth]{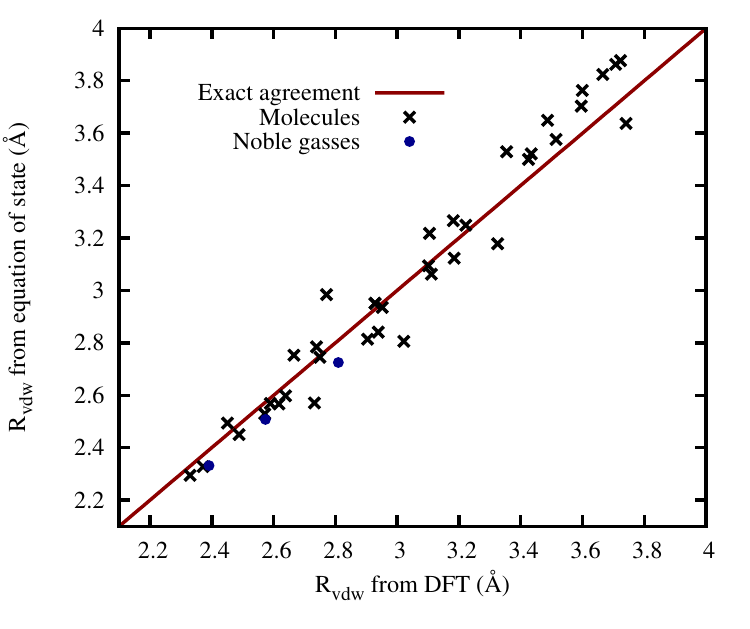}}
\caption{
    The van der Waals radius, calculated using DFT and using the equation of state for the gas phase.
    \label{fig:Rvdw}
}
\end{figure}
\end{center}

To determine the van der Waals radius ($R\sub{vdw}$), one could use 
the volume of exclusion in the van der Waals equation of state for the gas phase,
but such data is not available for all solvents.
Instead, we define a DFT volume of exclusion 
\begin{equation}V \equiv \int (1-s) \, d^3r \equiv (4\pi/3) R\sub{vdw}^3 \end{equation}
using the cavity shape function $s\left(\vec{r}\right)$, but now with $n_c$ set to $n_{vdW} = \sci{1.83}{-4}$ bohr\super{-3},
which we obtained by fitting to van der Waals radii which are available in the literature\cite{Rvdw1, Rvdw2}.
The results, which show good agreement with literature, are given in figure \ref{fig:Rvdw}.

\begin{center}
\begin{figure}
\center{\includegraphics[width=1\columnwidth]{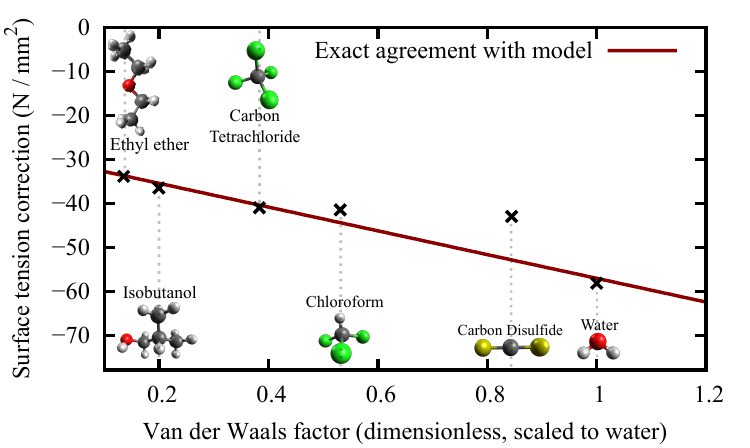}}
\caption{
    Surface tension correction ($\tau' = \tau\sub{eff} - \tau\sub{bulk}$)
    as a function of Van der Waals factor $N\sub{b} C_{solv} / R_{vdw}^5$ (scaled to that of water 
    to provide a dimensionless quantity for display purposes): data from fits to experimental solvation energies 
    \cite{truhlar, water-solvation-energies, water-solvation-energies-2, solvation-energies-3} (black x's), 
    best-fit linear correlation in the form of equation \ref{vdw} (red line).
    \label{fig:GammaFit}
}
\end{figure}
\end{center}

\begin{table*}
\small
\begin{tabular}{l | c c c | c c c}
\tiny
     & \multicolumn{3}{c}{\textbf{Best fit}} & \multicolumn{3}{c}{\textbf{Vapor constrained}}\tabularnewline

    \hline 
    \hline

    \multirow{2}{*}{\textbf{Solvents}} & $n_c$  & $\tau\sub{eff}$  & RMS Error & $n_c$ & $\tau\sub{eff}$ & RMS Error \tabularnewline
     & ($a_{0}^{-3}$) & (E\sub{H}/$a_{0}^{-2}$) & (kcal/mol) & ($a_{0}^{-2}$) & (E\sub{H}/$a_{0}^{-2}$) & (kcal/mol) \tabularnewline

    \hline
 
    Water & \sci{1.0}{-3} & \sci{\hskip10pt 9.50}{-6} & 0.95 & \sci{9.0}{-4} & \sci{\hskip10pt 1.02}{-5} & 1.38 \tabularnewline
    Chloroform & \sci{2.4}{-5} & \sci{-9.23}{-6} & 0.82 & \sci{5.6}{-4} & \sci{-1.11}{-5} & 1.30 \tabularnewline
    Carbon tetrachloride & \sci{1.2}{-4} & \sci{-8.99}{-6} & 1.02  & \sci{2.9}{-4} & \sci{-8.61}{-6} & 1.21 \tabularnewline
    Isobutanol & \sci{1.5}{-3} & \sci{-8.96}{-6} & 0.76 & \sci{1.8}{-3} & \sci{-8.28}{-6} & 0.83 \tabularnewline
    Carbon disulfide & \sci{2.9}{-5} & \sci{-7.96}{-6} & 1.01 & \sci{3.4}{-4} & \sci{-1.32}{-5} & 2.32 \tabularnewline
    Ethyl ether & \sci{2.6}{-4} & \sci{-1.08}{-5} & 1.13 & \sci{5.0}{-4} & \sci{-1.12}{-5} & 1.35 \tabularnewline

    \hline
       & & Average of RMS errors: & 0.95 & & Average of RMS errors: & 1.40 \tabularnewline
    \hline
    \multicolumn{7}{c}{$\gamma\sub{0}$ = \sci{-1.927}{-5} E\sub{H}/$a_{0}^{-2}$ \hskip25pt $\gamma\sub{1}$ = \sci{-1.313}{-2} } \tabularnewline

    \hline
\end{tabular}
\caption{
    PCM parameters and RMS errors for the solvents used in the determination of
    the values for $\gamma\sub{0}$ and $\gamma\sub{1}$.
    \label{tab:fitResults}
}
\end{table*}

With all of the parameters in equation \ref{vdw} defined, we next test our correlation hypothesis.
We begin by employing the standerd technique of deriving solvation model parameters from fits to solvation databases.
These fits allow us to determine the effective surface tension ($\tau\sub{eff}$)
for each of the six solvents in figure \ref{fig:GammaFit}.

Figure \ref{fig:GammaFit} shows that there indeed is a strong correlation between the correction term $\tau' = \tau\sub{eff} - \tau\sub{bulk}$
and our measure of dispersion strength $N\sub{b} C_{solv}/R_{vdw}^5$. 
The only apparent outlier in the fit set is CS$_2$, whose anomolous behavior we suspect is related to its being the only molecule in the fit set which has no net dipole moment while simultaneously having a significant non-zero quadrupole moment.
The case of ethylene glycol, another solvent with the same characteristic, is discussed later in the paper.

Based on the above observations, we propose as an approximate universal form for the effective cavity tension
\begin{eqnarray}
    \tau\sub{eff} & = & \tau\sub{bulk} + \gamma\sub{0} + \gamma\sub{1} \left[ \frac{N\sub{b} C_{solv} }{R_{vdw}^5} \right]
    \label{eqn:tEff},
\end{eqnarray}
where the first term $\tau\sub{bulk}$ is the bulk surface tension of the solvent
and is a measure of the energy cost to form macroscopic cavities in the liquid, the second term
$\gamma\sub{0} \equiv \sci{-1.927}{-5}$ E\sub{H}/$a_{0}^{-2}$ is a microscopic correction corresponding to the vertical intercept of the linear correlation, and the final term ($\gamma\sub{1} \equiv \sci{-1.313}{-2}$) incorporates the effects of long-range dispersion as the slope of the correlation.
Figure \ref{fig:GammaFit} illustrates our best fit values for $\gamma\sub{0}$ and $\gamma\sub{1}$ and compares the resulting linear model values for $\tau\sub{eff}$ with those which came from the original data-set fits, showing that we indeed can predict quite well appropriate values for this parameter without additional fiting to solvation data whatsoever.

Regarding the magnitude of our slope fit parameter $\gamma_1$, comparing equations \ref{vdw} and \ref{eqn:tEff},
we are able to extract from our fit a measure of the Grimme van der Waals scale factor, $s_6=32\times 3 \times \gamma_1=1.26$.  To place this value for $s_6$ in context, we note that, in the Grimme framework\cite{Grimme}, this parameter is generally fit to account first for the fact that some of the dispersion interaction (the short-range part) is accounted in standard approximate exchange-correlation functionals and, second, for the fact that a pair-potential model for the van der Waals interaction misses multiple molecule interactions.  In theory, the $s\sub{6}$ parameter would have a value of unity, but is known to change by as much as 45 \% between different electronic exchange-correlation functionals\cite{Grimme}, placing our fit result squarely in the expected range.

In addition to the readily available bulk surface tension $\tau\sub{bulk}$, 
only three solvent-dependent quantities are required to determine $\tau\sub{eff}$, namely
the bulk number density of the solvent $N\sub{b}$ (readily available from bulk thermodynamic data), 
the effective Van der Waals coefficient $C_{solv}$ (computed by summing the readily available and tabulated \cite{Grimme} atomic static dipole polarizabilities), and the effective Van der Waals radius of the solvent $R_{vdw}$,
which can be obtained using simple \emph{ab initio} calculations as described above.

With the non-electrostatic contributions now determined, we need only to define the electrostatic contibutions to complete our model.  
For these electrostatic interactions, we employ the non-linear dielectric response model of Gunceler et al\cite{NonlinearPCM}.  
In addition to n\sub{c}, this model requires solvated dipole moments, which we have determined self-consistently within our model fluids
using the procedure outlined in the same paper \cite{NonlinearPCM}. 
The numerical results for these dipole moments are given in table \ref{tab:sc-dipoles}.
Finally, to determine n\sub{c}, rather than employing a database of solvation energies,
we fit to a single datum, the self-solvation energy,
which can be easily determined from the vapor pressure\cite{Pvap}, which is more readily available. 
The resulting numerical values for n\sub{c} for the six solvents in our training set are reported in table \ref{tab:fitResults};
whereas the values for additional solvents of technological importance, are reported in table \ref{tab:more-solvents}.

\begin{table*}
\small
\begin{tabular}{l c c c}

    \textbf{Solvents} & $n_c$ ($a_{0}^{-3}$)  & $\tau\sub{eff}$ (E\sub{H}/$a_{0}^{-2}$) & RMS Error (kcal/mol) \tabularnewline
    \hline 
    \hline

    Acetone    & \sci{8.6}{-5} & \sci{-4.91}{-6} & \tabularnewline
    Acetonitrile & \sci{1.8}{-4} & \sci{-6.29}{-7} & \tabularnewline
    Dichloromethane & \sci{9.3}{-4} & \sci{-2.74}{-6} & 0.97 \tabularnewline
    Dimethyl sulfoxide & \sci{9.5}{-4} & \sci{\hskip10pt 8.42}{-6} & 2.09 \tabularnewline
    Ethylene carbonate & \sci{1.8}{-3} & \sci{\hskip10pt 1.55}{-5} & \tabularnewline
    Ethanol & \sci{1.3}{-3} & \sci{-5.10}{-6} & 1.40 \tabularnewline
    Glyme & \sci{8.3}{-5} & \sci{-8.03}{-6} & \tabularnewline
    Methanol    & \sci{6.5}{-4} & \sci{-5.23}{-6} & \tabularnewline
    Propylene Carbonate & \sci{9.8}{-4} & \sci{\hskip10pt 9.53}{-6} & \tabularnewline
    Tetrahydrofuran    & \sci{1.6}{-3} & \sci{-1.69}{-6} & 1.04 \tabularnewline
    Ethylene Glycol & \sci{5.4}{-4} & \sci{\hskip9pt 1.15}{-5} & see next section \tabularnewline
\end{tabular}
\caption{
    Parameters for solvents that were not used in the construction of the model.
    \label{tab:more-solvents}
}
\end{table*}

\begin{center}
\begin{table*}
\begin{tabular}{l c c c c}

           & \textbf{Water} & \textbf{Chloroform} & \textbf{Carbon tetrachloride} & \textbf{Isobutanol} \tabularnewline
    \hline
    Vacuum & 0.727          & 0.442               & 0.000                         & 0.627            \tabularnewline
    Liquid & 0.940          & 0.491               & 0.000                         & 0.646            \tabularnewline

           & \textbf{Carbon disulfide} & \textbf{Ethyl ether} & \textbf{Acetone} & \textbf{Dichloromethane} \tabularnewline
    \hline
    Vacuum & 0.0           & 0.409                & 1.185                         & 0.676 \tabularnewline
    Liquid & 0.0           & 0.487                & 1.387                         & 0.890 \tabularnewline

           & \textbf{Ethylene Carbonate} & \textbf{Glyme} & \textbf{Methanol} & \textbf{Tetrahydrofuran} \tabularnewline
    \hline
    Vacuum & 1.929         & 0.000                & 0.649                        & 0.720 \tabularnewline
    Liquid & 2.674         & 0.000                & 0.791                        & 0.909 \tabularnewline

           & \textbf{Acetonitrile} & \textbf{Dimethyl sulfoxide} & \textbf{Ethanol} & \tabularnewline
    \hline
    Vacuum & 1.581          & 1.606               & 0.604 & \tabularnewline
    Liquid & 1.892          & 2.192               & 0.762 & \tabularnewline

\end{tabular}
\caption{
    Effective dipole moment of solvents in liquid phase, calculated self-consistently.
    All are in atomic units ($e a_0$)
    \label{tab:sc-dipoles}
}
\end{table*}
\end{center}

\begin{figure}
    \includegraphics[width=1\columnwidth]{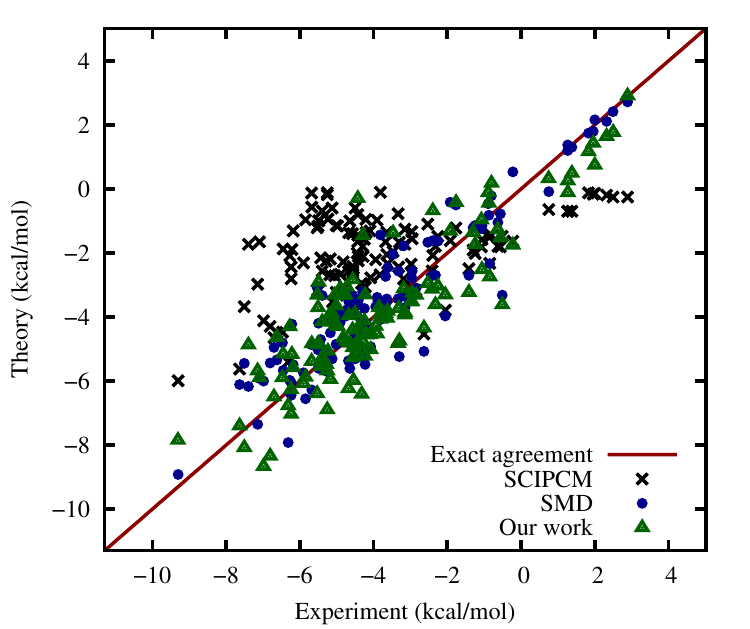}
\caption{
    Experimental and theoretical solvation energies for 124 solutes in 10 solvents computed in three theories: 
    SMD,\cite{SMD} SCIPCM,\cite{IPCM} and our work.
    \label{fig:compare}
}
\end{figure}

In figure \ref{fig:compare}, calculated solvation energies for 10 solvents,
\footnote{
The 10 solvents are: Carbon tetrachloride, Chloroform, Carbon disulfide, Dimethyl sulfoxide, dichloromethane, ethanol, ethylene glycol,
ethyl-ether, isobutanol and tetrahydrofuran.
}
of which 4 were not part of the fitting set,
are compared with two other solvation models available in the literature.
RMS errors in our work are between 0.8-1.4 kcal/mol for most solvents considered, 
except for two pathological cases that have errors greater than 2.0 kcal/mol.
The coefficient of determination, a standard measure of predictive power, is 0.76.
Solvation energies calculated with our approach are competitive with but somewhat worse than SMD, \cite{SMD}
a commonly used and very successful solvation model that uses the union-of-spheres approach for determining cavities.
(SMD has RMS errors between 0.6-1.7 kcal/mol and a coefficient of determination of 0.83
for the same set of solvents and solutes.)
An interesting observation is that even though both theories tend to undersolvate,
SMD has a slightly larger bias (mean error 0.296 kcal/mol) than our work (mean error 0.027 kcal/mol), 
which might be due to the large number of polar solutes in our training set.
SCIPCM, \cite{IPCM} another iso-density model that shares some traits with our work, is not competitive for non-aqueous solvents 
and has a negative coefficient of determination for them.
We believe that this is due to the insufficient accounting of non-electrostatic effects in SCIPCM.
These results are very encouraging because, to our knowledge, this is the first attempt to universalize iso-density PCMs 
whereas SMD, and other universal models from the same tradition, 
have almost two decades of research and optimization behind them. \cite{SMx96, SMxReview, SMD}

Despite the apparent success of the iso-density approach, there are several important inadequacies
which require further work to overcome.
The most of important, in our opinion, is the fact that the dispersion interaction 
is treated only at an effective surface tension level.
This makes it difficult for the theory to distinguish between similarly sized solutes
if the electrostatic interaction is very weak.
This is not a problem for polar solutes (such as alcohols or thiols), but may be a problem for some less polar ones.
For example, this theory would predict very similar solvation energies for hydrocarbons and their corresponding fluorocarbons,
when in fact, the solvation energies might be very different.

Furthermore, there are also inadequecies resulting from the underlying electrostatic model.
In this work, we used the nonlinear continuum model by \textcite{NonlinearPCM}.
This model correctly captures nonlinear dielectric response resulting from the rotational saturation of permanent dipoles in the solvent.
However, problems arise if the solvent molecule has no dipole moment, but has a significant nonzero quadrupole moment.
One example of is was CS\sub{2}, but an even more extreme example would be ethylene gylchol.
The problem arises because ethylene glycol is 
essentially two dipolar units (each resembling a methanol molecule) attached together.
The interaction between the solute and solvent depends very strongly on the orientation of these dipole groups near the solute.
This dependence is correctly captured for most other solutes, but not for solvents like ethylene glycol as the net dipole moment is zero.
To explore this issue, we do the following: 
Instead of using the overall dipole density of the molecule (which is zero),
we use twice the dipole density of its constituent pieces, which in this case is that of methanol.
As seen in table \ref{tab:ethylene-glycol}, this procedure improves solvation energies,
indicating that the quadrupole moment is indeed the source of the problem.
We believe that more sophisticated electrostatics models, such as those using non-local response, \cite{candle}
might have greater success in these pathological cases.

\begin{table*}
\begin{tabular}{l | c c >{\bfseries}c | c >{\bfseries}c}
    \textbf{Solute} & Experiment & Model & Error & Model (corrected) & Error (corrected) \tabularnewline
    \hline 
    \hline
    benzene       &  0.83 & -4.41 & -5.24 & 1.41 & 0.56 \tabularnewline
    chlorobenzene & -0.26 & -5.54 & -5.28 & 1.52 & 1.78 \tabularnewline
    fluorobenzene &  0.69 & -4.84 & -5.53 & 1.09 & 0.40 \tabularnewline
    naphthalene   & -2.15 & -5.91 & -3.76 & 1.80 & 3.95 \tabularnewline
    toluene       &  0.74 & -4.83 & -5.57 & 1.93 & 1.19
\end{tabular}
\caption{
    Model predictiontions and experimental values for the solvation energy in ethylene glycol.
    All values are in kcal/mol.
    \label{tab:ethylene-glycol}
}
\end{table*}

{\em Conclusion ---} 
In conclusion, this work presents a universal isodensity solvation model for ab-initio calculations in a wide range of polar and nonpolar solvents based only on readily obtainable bulk thermodynamic data
and {\em ab initio} computables, {\em without the need for a database of solvation energies} to fit the model parameters for each new solvent of interest.
This work thus opens to investigation a wide range of solvents previously inaccessible to iso-density solvation studies, opening new application areas, in particular those at solid-liquid  interfaces, to plane-wave {\em ab initio} study.

{\em Computational details ---} 
We performed all plane-wave calculations with JDFTx \cite{JDFTx}, 
an open-source implementation of joint density-functional theory.
We employed the revTPSS meta-gga approximation \cite{revTPSS} for the electronic exchange-correlation
and norm-conserving pseudopotentials generated using the Opium pseudopotential generation package\cite{opium} to represent the ionic cores.
Kohn-Sham orbitals are expanded using planewaves up to a cutoff of 30 Hartrees.  
We obtained molecular geometries from the CCCBDB database\cite{cccbdb}.
For solvation energies in SMD \cite{SMD} and SCIPCM \cite{IPCM} models, we used a 6-31G* basis set.

\begin{acknowledgements}
The authors would like to thank Ravishankar Sundararaman, Yalcin Ozhabes and Prof. Ersen Mete for stimulating discussions.
This work was supported as a part of the Energy Materials Center at Cornell (EMC\super{2}),
an Energy Frontier Research Center funded by the U.S. Department of Energy, Office
of Science, Office of Basic Energy Sciences under Award Number DE-SC0001086.
\end{acknowledgements}

\bibliography{references}

\end{document}